\documentclass[aps,prl,twocolumn,superscriptaddress,showkeys]{revtex4-2}

\usepackage{graphicx}

\usepackage{pdfpages}
\usepackage{pgffor}

\makeatletter
\AtBeginDocument{\let\LS@rot\@undefined}
\makeatother

\begin{document}

\title{Improper-proper ferroelectric competition as a mechanism for multistate polarisation and ferrielectric-like behaviour}

\author{Cameron A.M. Scott}

\affiliation{Centre for Materials Physics, Durham University, South Road, Durham DH1 3LE, United Kingdom}
\affiliation{Luxembourg Institute of Science and Technology (LIST), Avenue des Hauts-Fourneaux 5, Esch-sur-Alzette, L-4362, Luxembourg}
\author{Finlay D. Morrison}

\affiliation{EaStCHEM School Of Chemistry, University Of St Andrews, St Andrews, KY16 9ST, United Kingdom}

\author{Nicholas C. Bristowe}

\affiliation{Centre for Materials Physics, Durham University, South Road, Durham DH1 3LE, United Kingdom}


\begin{abstract}
In this paper, we re-explore a simple textbook Landau model describing improper ferroelectricity and show that in the limit where both proper and improper instabilities exist and compete, improper ferroelectrics can display switching between multiple polarisation states. Using first principles calculations we highlight how the hexagonal tungsten bronze materials may be an archetypal case, with the possibility to switch between improper and proper phases. The resulting functional characteristics are akin to ‘ferrielectrics’, with switching behaviour in the form of a triple hysteresis loop. Such functionality could be ideal for creating non-volatile multistate systems for use in memory devices or as a backbone for neuromorphic computing. 
\end{abstract}

\maketitle

\textit{Introduction} - Ferroic phase transitions are characterised by the appearance of a spontaneous and switchable order parameter (OP) as a material passes from a high-symmetry to a low symmetry phase. In ferroelectric phase transitions \cite{rabe2007physics,tagantsev2010domains}, the OP is the net electrical polarisation. However, the polarisation may not be the only OP involved in the phase transition and ferroelectrics can be further classified by considering the role of other OPs. In proper ferroelectrics such as the perovskite PbTiO$_3$, the polar mode in the high-symmetry structure softens with decreasing temperature, eventually becoming unstable at the transition temperature, resulting in a spontaneous polar distortion \cite{garcia1996first}.  Due to couplings to the polarisation, other OPs may appear below the transition temperature, but the driving distortion is polar. In contrast, improper ferroelectrics \cite{levanyuk1974improper} are materials in which the high-symmetry phase possesses a softening non-polar distortion which becomes unstable at the transition temperature. Owing to odd-order couplings between OPs, a polarisation appears even though only the non-polar modes are unstable. Improper ferroelectrics have recently received a great deal of attention because they circumvent the narrow requirements that typically promote proper ferroelectricity \cite{hill2000there,benedek2013there}. Non-polar distortions are largely ubiquitous and so improper ferroelectricity can be engineered in a much broader range of materials. Furthermore, the non-polar modes that produce an electrical polarisation, which can include magnetic orderings or antiferrodistortive motions of cations, can be controlled through a variety of external stimuli, leading to the exciting situation where the electric polarisation can be controlled using either magnetic fields or strain \cite{benedek2011hybrid,schlom2007strain}.

The Landau theory of phase transitions is a completely general methodology for studying both proper and improper ferroelectric phase transitions \cite{toledano1987landau}. In this theory, the free energy of the system is expanded as a polynomial of the order parameters. The polynomial is determined by the requirement that each term be invariant under all the symmetry operations of the high-symmetry phase; the minimum of the free energy then determines the state of the system. A simple model, commonly found in textbooks \cite{toledano1987landau,cowley1980structural}, that can describe both proper and improper phase transitions involving two order parameters (a polar mode $\Gamma$ and a non-polar mode $K$) would have a free energy of the form

\begin{equation}
\label{eq:landscape}
\mathcal{F}(\Gamma,K) -\mathcal{F}_0= a_\Gamma \Gamma^2 + b_\Gamma \Gamma^4 + a_K K^2 + b_K K^4 + \beta \Gamma^2 K^2 + \lambda \Gamma K^3,
\end{equation}

relative to the energy $\mathcal{F}_0$ of the high symmetry phase. The fourth order coefficients $b_\Gamma$ and $b_K$ must be positive, because Equation \ref{eq:landscape} must have a global minima for finite values of the order parameters. As it is always allowed at fourth-order, we have also included a biquadratic coupling term. It is known that polar and non-polar modes typically compete at the biquadratic level \cite{benedek2013there,scott2024universal}, so that $\beta$ is also positive. With these conditions, it is quite evident that for $a_\Gamma, a_K > 0$, the free energy has a minimum when $\Gamma,K = 0$.  Ignoring the final linear-cubic coupling term for now, then if $a_\Gamma < 0$ but $a_K > 0$, the system is minimised by $\Gamma \neq 0$ and $K = 0$. Therefore, the sign of the quadratic coefficients indicates the stability of the corresponding OP and in this case, only the polar OP $\Gamma$ is unstable, suggesting proper ferroelectricity. Improper ferroelectrics can be studied by including the final coupling and considering a situation where $a_K < 0$ but $a_\Gamma > 0$. Only $K$ is unstable, but the final term guarantees that the energy is minimised with both $\Gamma, K \neq 0$.

Improper ferroelectricity of this kind is typified by hexagonal YMnO$_3$, which experiences a series of phase transitions as the temperature is decreased \cite{gibbs2011high,abrahams2001ferroelectricity,van2001hexagonal}. The first of these, occurring between 1243 and 1273 K and reducing the symmetry from the centrosymmetric $P6_3/mcm$ to the polar $P6_3cm$, is crucial for establishing the improper nature of this ferroelectric transition \cite{levanyuk1974improper,dvovrak1974improper,toledano1987landau}.  At this transition, the unit cell triples in size and zone-boundary ($k = (1/3,1/3,0)$) non-polar and zone-centre ($k= (0,0,0)$) polar distortions appear as OPs. These two OPs are labelled with their respective irreducible representations $K_3$ and $\Gamma_2^-$, in the notation of Miller and Love \cite{cracknell1979kronecker}. Using the rules of invariant analysis \cite{senn2018group} and the symmetry of the two phases, it is possible to write the free energy in the form of Equation \ref{eq:landscape}. Computational studies \cite{van2004origin,fennie2005ferroelectric} confirmed that $a_\Gamma > 0$ whilst $a_K < 0$, indicating that it is the non-polar mode that is unstable. This result unambiguously confirmed that hexagonal YMnO$_3$ was an improper ferroelectric. 

The distinction between proper and improper ferroelectricity is firmly established in the literature - many textbooks now provide detailed discussions \cite{lines2001principles,khomskii2010basic} - although recent studies in perovskite superlattices, as well as in Ruddlesden-Popper and Dion-Jacobson phases, have revealed shallow energy landscapes and subsequent competition between \textit{hybrid-improper} ferroelectricity and either proper ferroelectricity or antiferroelectricity, so that the distinction becomes less clear \cite{cascos2020tuning,mallick2021switching,yoshida2018ferroelectric,benedek2022hybrid,gou2014piezoelectricity}. Possibly due to the lack of demonstrative materials, the analogous competition between proper and improper ferroelectricity, for which $a_\Gamma, a_K < 0$ serves as a typical example, remains largely unexplored. 

In this paper, we study this situation and find that this small alteration to the textbook model results in dramatic changes to the energy landscape and enables switching between proper and improper ferroelectricity. To the best of our knowledge, the present study is the first to identify a physical realisation of the extended model. Specifically, we focus on the hexagonal tungsten bronze CsNbW$_2$O$_9$, which was recently synthesised by McNulty \textit{et al} \cite{mcnulty2019electronically} and, due to symmetry, has an energy functional described by Equation \ref{eq:landscape}, but with $a_\Gamma, a_K < 0$. Using density functional theory calculations, we confirm that only $a_K$ is negative in YMnO$_3$, but that both $a_K$ and $a_\Gamma$ are negative in CsNbW$_2$O$_9$, which creates a complex energy landscape with four near-degenerate minima. We find that if the system exists in two of these minima, its behaviour should resemble the improper behaviour of YMnO$_3$ and if the state exists in the other two minima, the system would behave like a proper ferroelectric. Furthermore, we include simple temperature and electric field dependences to our free energy polynomial and conclude that either a reduction in temperature or the application of an external electric field may be enough to change the nature of ferroelectricity in CsNbW$_2$O$_9$. As our results are based on considerations of symmetry, our theory is general and should hold in any improper ferroelectric which also has an independent polar instability and we propose that materials with these properties, such as the hexagonal tungsten bronzes, should be studied to explore the competition between proper and improper mechanisms. Finally, we argue that in instances with such competition, the functional behaviour of the system is akin to ferrielectricity and could display complex hysteresis loops for multistate memories.

\textit{Computational Details} -We performed quantum mechanical simulations using density functional theory (DFT) as implemented in the Vienna Ab-Initio Software Suite (VASP) \cite{kresse1996efficient,kresse1996efficiency,kresse1999ultrasoft} with the PBESol exchange-correlation functional \cite{perdew2008restoring}. All calculations were performed in cells large enough to accommodate the cell tripling distortion. To obtain a convergence of the energies and forces in YMnO$_3$ of 1 m$e$V/f.u and 1 m$e$V/{\AA} respectively, we employed a 600 $e$V plane-wave cutoff and a $k$-grid of 5x5x2. For the same convergence in CsNbW$_2$O$_9$, an 800 $e$V cutoff and a $k$-grid of 3x3x9 were needed. When the low temperature phase including tilts is studied \cite{mcnulty2019electronically}, a larger cell is required and so a 3x3x4 grid is used instead. Modelling the disorder on the Nb/W sites was performed using the virtual crystal approximation (VCA) in which the pseudopotentials of each atom are linearly mixed in proportion to their stoichiometry \cite{bellaiche2000virtual}. To test the feasibility of this approach, we computed the energies of the four known crystal structures of CsNbW$_2$O$_9$ using VCA and find that the energies corroborate the observed sequence of phase transitions in the material - see Table S1. All symmetry analysis and decomposition of structures into symmetry-adapted modes was performed using the ISOTROPY Software Suite \cite{campbell2006isodisplace,hatch2003invariants}.

\begin{figure*}
\includegraphics[width=0.8\textwidth]{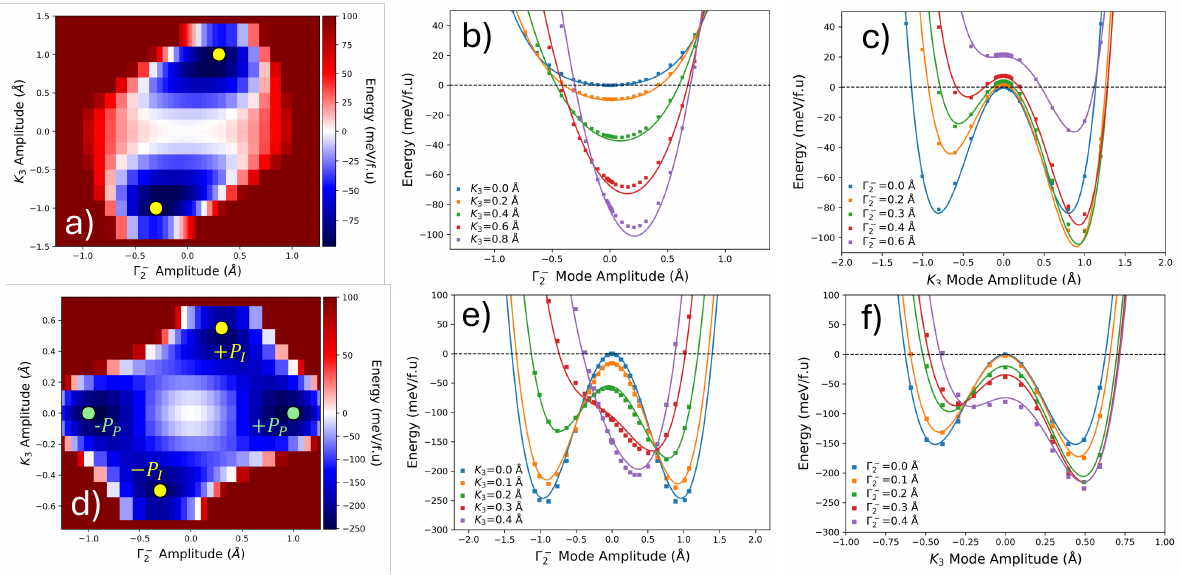}
\caption{\label{landscape}Energy landscapes of a)YMnO$_3$ and d) CsNbW$_2$O$_9$ described by Equation \ref{eq:landscape}.  Yellow points denote the minima in the landscape with improper character. Green points denote the minima with proper character. The minima of the free energy landscape in Figure 1d have been labelled with the polarisation of each minimum, so that the proper-like and improper-like minima have polarisation $\pm P_P$ and $\pm P_I$ respectively. Panels b) c) e) and f) explore one dimensional slices of these landscapes and illustrate the contrast between the canonical improper behaviour of YMnO$_3$ and the novel behaviour of CsNbW$_2$O$_9$. Sixth-order fits have been used here for illustrative purposes, but a simpler fourth order fitting is used to extract the coefficients in Table \ref{tab:coefficient_wells} - a comparison is made in the SI. }
\end{figure*}

\textit{Results And Discussion} - To explore the energy landscapes of YMnO$_3$ and CsNbW$_2$O$_9$, we performed a full geometry relaxation of the high-symmetry phases of each. We then use the ISODISTORT tool of the ISOTROPY Suite to create distorted structures with varying amplitudes of the $\Gamma_2^-$ and $K_3$ modes. Figure \ref{landscape}a plots the two-dimensional energy landscape of YMnO$_3$. The yellow points mark the two symmetry equivalent minima that are obtained by simultaneously introducing both $\Gamma_2^-$ and $K_3$ modes. Figure \ref{landscape}b is a one-dimensional slice through this landscape, illustrating how the energy varies upon changing the magnitude of the $\Gamma_2^-$ mode with fixed $K_3$. We observe the canonical behaviour of an improper ferroelectric wherein the polar $\Gamma_2^-$ mode is a single well with positive curvature and a minimum that shifts away from $\Gamma_2^- = 0$ with increasing $K_3$. Figure \ref{landscape}c shows the equivalent well for $K_3$, displaying a transformation of a symmetric double well potential (indicating an instability in $K_3$) into a highly asymmetric well. From Figure \ref{landscape}a, we can extract the coefficients of Equation \ref{landscape}. These are tabulated in Table \ref{tab:coefficient_wells}. This result confirms the conclusion of Reference \cite{fennie2005ferroelectric} - YMnO$_3$ is an improper ferroelectric with an instability in the $K_3$ mode and the improper coupling $\lambda \Gamma K^3 $ drives the ferroelectric polarisation.

\begin{figure}
\includegraphics[width=0.8\columnwidth]{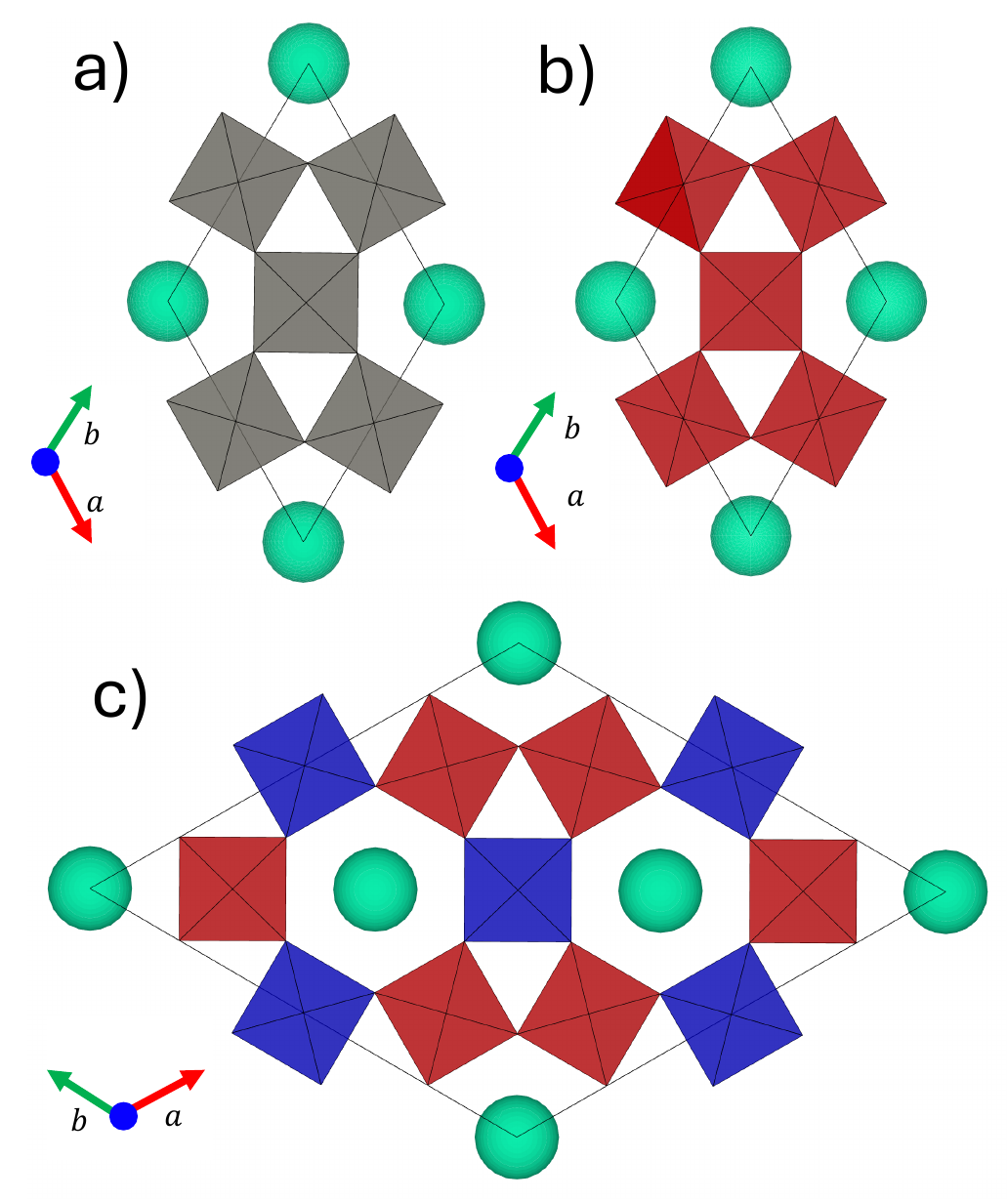}
\caption{\label{structure}Representation of the crystal structures of CsNbW$_2$O$_9$.  Green spheres are Cs atoms. Red and blue oxygen octahedra indicate that the central (Nb/W) atom has shifted up or down along the polar $c$ axis, respectively. a) Highest symmetry structure with no SOJT displacements in any octahedra. b) Crystal structure obtained in the proper-like minima, where all octahedra have a coordinated displacement of their central atom along the polar axis. c) Crystal structure of the improper-like minima involved an $\uparrow\downarrow\uparrow$ pattern of displacements with the associated cell tripling. }
\end{figure}

CsNbW$_2$O$_9$ is a similar material in the sense that there is a high temperature centrosymmetric phase with $P6/mmm$ symmetry (Figure \ref{structure}a) which transitions to a tripled non-centrosymmetric structure with $P6mm$ symmetry upon a reduction in temperature (Figure \ref{structure}b). Like YMnO$_3$, this transition is characterised through the $K_3$ and $\Gamma_2^-$ irreducible representations at the $k = (1/3,1/3,0)$ and $k= (0,0,0)$ points in the Brillioun zone, respectively. Due to this similarity, the material can again be described by a free energy expansion identical to Equation \ref{eq:landscape}. However, the energy landscape of the hexagonal tungsten bronze CsNbW$_2$O$_9$, shown in Figure \ref{landscape}d, contains four minima. Two of these, with non-zero $\Gamma_2^-$ and $K_3$, are analogous to the improper phase of YMnO$_3$. The other two minima, marked in green, are unique to the hexagonal tungsten bronze and describe a phase in which the antipolar $K_3$ mode has vanished but the polar $\Gamma_2^-$ mode has a substantially larger amplitude than in the improper-like minimum. We refer to these novel minima as proper-like and the corresponding distortions are shown in Figure \ref{structure}b. The one-dimensional slices in Figure \ref{landscape}e and \ref{landscape}f  reveal that CsNbW$_2$O$_9$ is qualitatively different from a canonical improper ferroelectric. Indeed, the negative value of $a_\Gamma$ in Table \ref{tab:coefficient_wells} creates a double well that distorts into a single well with increasing $K_3$. We can also see from this landscape that the proper-like minimum is actually the global minimum, by approximately 20 m$e$V/f.u. As can be seen in the Supplemental Information, the barriers between the proper-like and improper-like minima are sizeable fractions of the well depths. This suggests that once the state enters one minima, it may well be kinetically trapped there, even if that particular minimum is not globally stable. We note that adding higher order terms to Equation \ref{eq:landscape} does not qualitatively change the fitted landscape; Figure S1 shows that four minima still exist but the relative energies slightly change.

\begin{table}[h]
\caption{\label{tab:coefficient_wells}Coefficients of Landau expansion in Equation \ref{eq:landscape} fitted to DFT.}
\begin{ruledtabular}
\begin{tabular}{lccc}
Coefficient & YMnO$_3$& CsNbW$_2$O$_9$  \\
\hline
$a_\Gamma$ (meV/$\mathrm{\AA}^2$ f.u)   & 26 $\pm$ 2  & -470 $\pm$ 10   \\
$a_K$ (meV/$\mathrm{\AA}^2$ f.u)   & -259 $\pm$ 2  & -1360 $\pm$ 20   \\
$b_\Gamma$ (meV/$\mathrm{\AA}^4$ f.u)   & 62 $\pm$ 2  & 260 $\pm$ 10   \\
$b_K$ (meV/$\mathrm{\AA}^4$ f.u)   & 200 $\pm$ 2  & 3490 $\pm$ 30  \\
$\beta$ (meV/$\mathrm{\AA}^4$ f.u)   & 537 $\pm$ 2  & 4280 $\pm$ 20   \\
$\lambda$ (meV/$\mathrm{\AA}^4$ f.u)   & -313 $\pm$ 1  & -3380 $\pm$ 10   \\
\end{tabular}
\end{ruledtabular}
\end{table}

Next, we performed geometry relaxations of the proper-like and the improper-like phases, allowing all atomic positions and lattice vectors to change until the energy is minimised. As detailed in Table S2, we can stabilise the two phases, demonstrating that the minima are robust to the effects of both strain and couplings to other secondary modes \cite{scott2024universal,benedek2013there}. This justifies our truncation of Equation \ref{eq:landscape} to fourth-order, as higher order terms or strain clearly do not destroy the pertinent features of the landscape. CsNbW$_2$O$_9$ undergoes further phase transitions at lower temperatures which introduce two tilt modes ($A_3^+$ and $A_6^+$) and a multitude of other secondary OPs. We included these two tilt modes to both the improper and proper phases, which lowers the symmetry further to $Cmc2_1$, and again allowed this structure to fully relax. As shown in Table S6, we found that there still exists a distinct proper and improper minimum even in the presence of these large tilt modes which are known to be competitive with polar distortions \cite{benedek2013there}. In all situations, the proper-like minimum remains lower in energy than the improper-like.  

Why is the energy landscape in CsNbW$_2$O$_9$ completely different to that in YMnO$_3$? In YMnO$_3$, the cell tripling mode $K_3$ corresponds to a buckling of the planes of O$_5$ tetrahedral bipyramids whilst the polar mode $\Gamma_2^-$ represents an off-centering of Y$^{3+}$ cations. In contrast, both the unit cell tripling distortion $K_3$ and the polar mode $\Gamma_2^-$ in CsNbW$_2$O$_9$ arise from of Nb$^{5+}$/W$^{6+}$ cations displacing from the centre of their octahedral environments due to a second order Jahn-Teller (SOJT) effect \cite{mcnulty2019electronically}; the former being a non-polar displacement pattern $\uparrow{\Big\downarrow}\uparrow$,  and the latter being a $\uparrow{\uparrow}\uparrow$ displacement pattern. In sharp distinction to the disparate modes in YMnO$_3$, the two modes in CsNbW$_2$O$_9$ are clearly related, and are likely to be in competition. To make this distinction quantitative, we calculated the Born effective charge tensor $Z_{ij}$ in the high-symmetry structures of YMnO$_3$ and CsNbW$_2$O$_9$, finding that the average effective charge along the polar axis ($Z_{zz}$) for each ion is Y$^{4.11+}$Mn$^{3.83+}$O$^{2.65-}_3$ and  Cs$^{1.26+}$Nb$^{11.29+}$W$^{11.39+}_2$O$^{3.92-}_9$. It is quite clear that the effective charge on the Nb and the W sites are anomalous, which confirms that the mechanism of ferroelectricity is driven by the SOJT effect.

Having established the existence of the proper-like minima, we explore mechanisms to cause transitions to them from the experimentally observed improper-like minima. To do this, we take the fitted parameters from Table \ref{tab:coefficient_wells} to construct the landscape of Equation $\ref{eq:landscape}$. Next, we introduce the textbook temperature dependence \cite{toledano1987landau} to the quadratic coefficient $a_\Gamma(T) = \frac{|a_\Gamma|}{T_C^\Gamma}(T-T_C^\Gamma)$ and $a_K(T) = \frac{|a_K|}{T_C^K}(T-T_C^K)$, defined in this way so that $a_\Gamma(0) = -|a_\Gamma|$ and $a_K(0) = -|a_K|$. With this definition, we have had to introduce two transition temperatures, $T_C^\Gamma$ and $T_C^K$, the former of which we treat as a free parameter. For $T_C^K$, we use the experimentally determined value of 1100 K. Using this fit, we observed four qualitatively different behaviours when we allow all the free parameters to vary by $\pm 5\%$,  which could be physically achieved through chemical substitution, epitaxial strain or pressure. Additionally, these changes in parameters are of the same order of magnitude to the error of the fit.

\begin{figure*}
\includegraphics[width=0.7\textwidth]{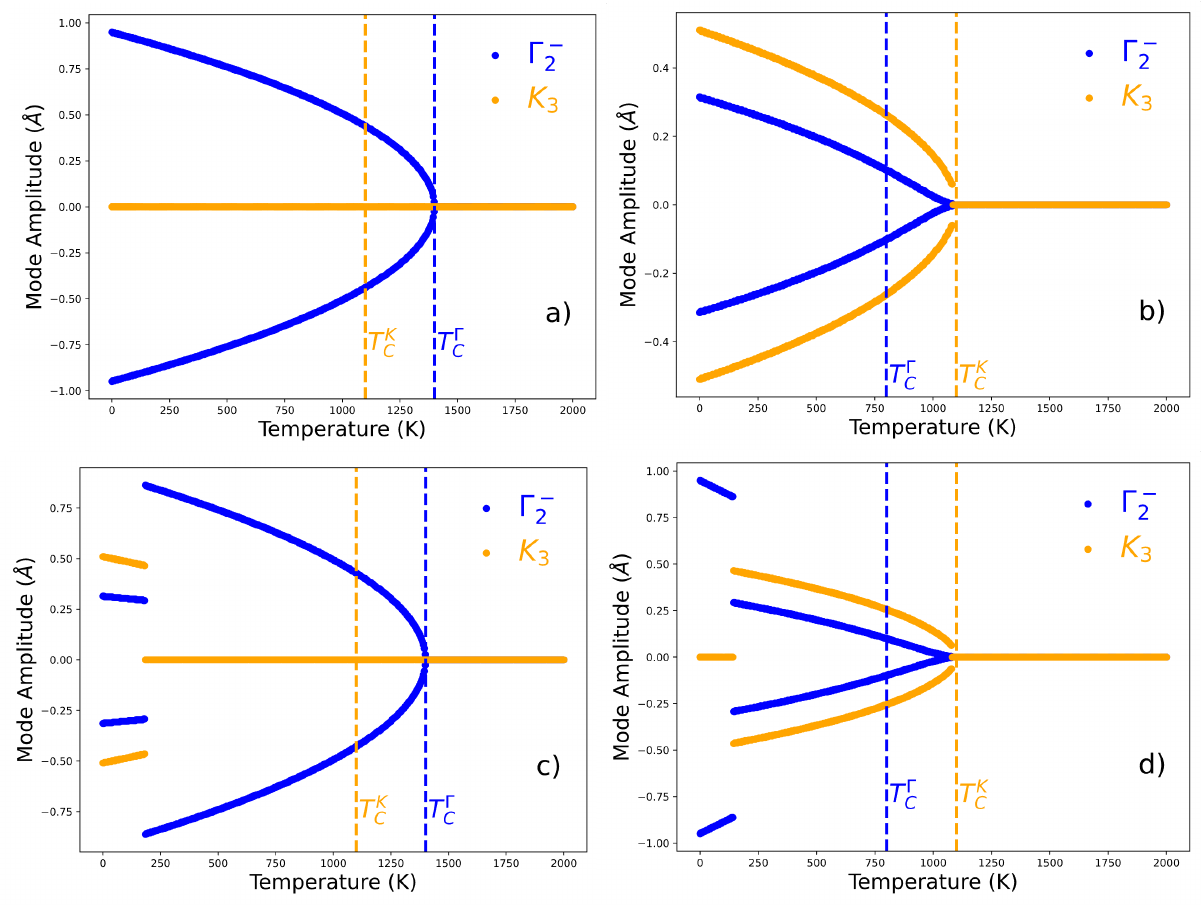}
\caption{\label{behaviours}Temperature dependence of the OPs. In each calculation we keep $\{b_\Gamma,b_K,\beta,\lambda\}=\{260,3490,4280,-3380\}$ and fix $T_C^K = 1100$ K, but change the other parameters so in a) we use $\{a_\Gamma,a_K,T_C^\Gamma\} = \{-470,-1360,1400\}$ and we observe proper-like behaviour. In b) we use $\{a_\Gamma,a_K,T_C^\Gamma\} = \{-445,-1430, 800\}$ and see improper-like behaviour. In c), we use $\{a_\Gamma,a_K,T_C^\Gamma,T_C^K\} = \{-445,-1430,1400\}$ and a proper-like to improper-like transition becomes possible. Finally, in d), we use $\{a_\Gamma,a_K,T_C^\Gamma,T_C^K\} = \{-470,-1360,800\}$ and there is an improper-like to proper-like transition. See Table \ref{tab:coefficient_wells} for the units of the parameters.}
\end{figure*}

Figure \ref{behaviours} plots the amplitudes of $\Gamma_2^-$ and $K_3$ in the global thermodynamic minimum of the free energy landscape. Due to the symmetry of Equation \ref{eq:landscape}, there are always two symmetrically equivalent minima. For the parameters of Figure \ref{behaviours}a, the proper minimum is always the global ground state and the $\Gamma_2^-$ amplitude exhibits the $\sqrt{T-T_C^\Gamma}$ dependence characteristic of second-order phase transitions within a mean-field theory \cite{goldenfeld2018lectures}. Despite $a_K$ being strongly negative, it is never enough to make an improper-like thermodynamic minimum. In short, this behaviour is indistinguishable from a proper ferroelectric like PbTiO$_3$ \cite{cohen1992electronic}. In Figure \ref{behaviours}b, the behaviour of the OPs looks exactly like a traditional improper ferroelectric like YMnO$_3$ \cite{lilienblum2015ferroelectricity}. The $K$ mode condenses and the improper coupling guarantees that the $\Gamma$ mode appears. The state persists in this minimum down to 0 K without any further phase transitions (tilts are not considered here). Just as Figure \ref{behaviours}a was indistinguishable from a proper ferroelectric, Figure \ref{behaviours}b is indistinguishable from an improper one. It should be kept in mind that because of the negative quadratic parameters $a_\Gamma$ and $a_K$, the proper-like minima still exist but are not globally stable. 

Figure \ref{behaviours}c shows qualitatively different behaviour. Now the $\Gamma_2^-$ mode appears first in a proper-like mechanism. However, the $K$ instability is relatively strong and so upon reducing the temperature, the improper-like minimum becomes globally stable and we observe a transition from a proper ferroelectric to an improper one. The parameters in Figure \ref{behaviours}d allow for the inverse of Figure \ref{behaviours}c in that the $K_3$ mode arrives first, bringing along the $\Gamma_2^-$ mode in an improper mechanism. However, the $\Gamma_2^-$ instability is now stronger and can stabilise the proper minimum. There is a transition from an improper ferroelectric to a proper ferroelectric.

For each of the behaviours, we also use Equation \ref{eq:landscape} to compute the energies of the minima, as well as the energy barriers between them. This is shown in Figures S2, S3, S4 and S5. Also computed here is the dielectric response that we predict for each type of phase transition using the fourth-order energy expansion of Equation \ref{eq:landscape}. The dielectric response helps to clarify the type of ferroelectric phase transition; proper transitions have diverging dielectric susceptibilities whilst improper transitions have only marginal changes in the electric response. We again note the large energy barriers between minima. These sizeable barriers would typically prohibit a transition between the two classes of minima at low temperatures, but may allow for one at elevated temperatures when the minima have only just formed.

How else is one to access these other minima? If we assume a simple, linear relationship between electrical polarisation $P$ and $\Gamma_2^-$ amplitude, such as $P = Z\Gamma$ with $Z$ being the charge moved by the distortion (equivalent to the average of the Born effective charges of all atoms moved by the $\Gamma$ mode \cite{king1993theory}), then an electric field can be added via a term of the form $-EZ\Gamma$. We set $Z=1 e$/f.u. for simplicity but it is trivial to amend with more realistic values. The effect of an electric field is easily understood as a switching between the minima of Figure \ref{landscape}d. For example, for a system trapped in the improper-like minimum labelled $+P_I$ in Figure \ref{landscape}d, a positive electric field will bias the state towards the minimum $+P_P$. However, the barriers between states are large, and so the state may be kinetically trapped in the improper minimum for fields well beyond that which makes the proper-like minimum thermodynamically stable. Fortunately, the electric field actually distorts the entire landscape, changing the relative energies between minima, altering barrier heights and even creating or destroying minima for large enough fields. This is depicted in Figure S6.

\begin{figure}
\includegraphics[width=0.8\columnwidth]{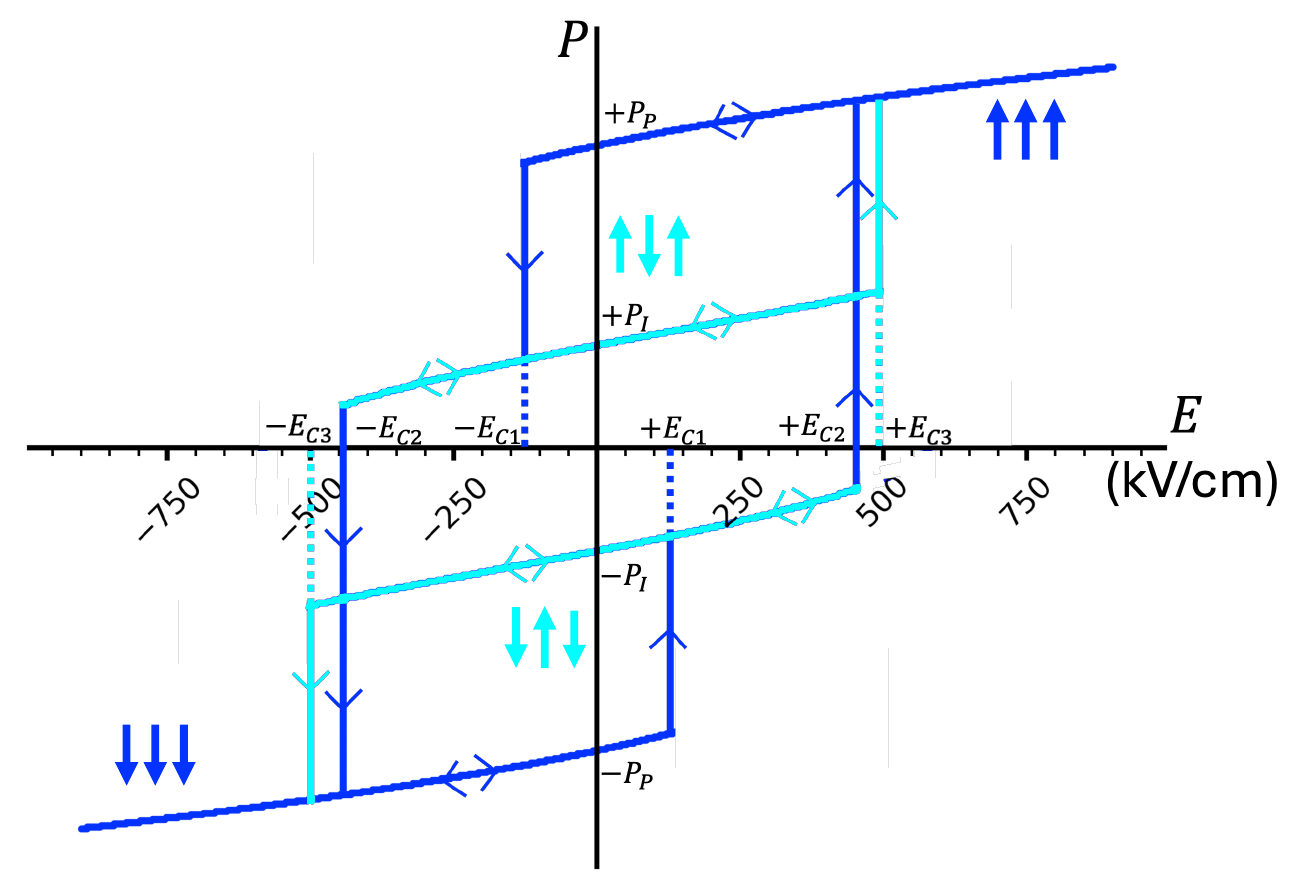}
\caption{\label{electric}Hysteresis loop for CsNbW$_2$O$_9$. This calculation was performed at $T=300$ K and $\{a_\Gamma,a_K,T_C^\Gamma \} =  \{-445,-1430, 800\}$. The coercive fields were estimated using the barriers in Figure S7. $\pm P_P$ and $\pm P_I$ refer to the polarisation of the proper-like and improper-like minima, respectively. The two colours are used to separate the transitions at $E_{C2}$ and $E_{C3}$ i.e. the system cannot transition from $+P_I$ to $+P_P$ at $E_{C2}$ but persists in the improper minimum until $E_{C3}$. The inset arrows give a schematic representation of the electric dipoles in each branch of the hysteresis loop.}
\end{figure}

In Figure \ref{electric}, we took the parameter set for Figure \ref{behaviours}b and swept an electric field for $T$ = 300 K, well below the transition temperatures. Considering the full range of electric fields, and using the barriers in between minima to calculate the three coercive fields $E_{C1}$, $E_{C2}$ and $E_{C3}$ (see Figure S7), we constructed a hysteresis loop. This is displayed in Figure \ref{electric} and we note that all four minima are accessible at zero field, so that hexagonal tungsten bronzes like CsNbW$_2$O$_9$ - or other materials with similar properties - could be the foundation of novel four-state non-volatile memory devices. Although Figure \ref{electric} resembles a triple hysteresis loop, we note that it is only possible to traverse the two outer loops in a straightforward manner. It is energetically unfavourable to go over the maximum at the origin of Figure \ref{landscape}d and so traversing the central loop is not possible. A direct transition from $+P_I$ to $-P_I$ with increasing negative field is prohibited, and so one must first transition to $-P_P$, reverse the direction of the field, and then to $-P_I$.  

Although the required coercive fields seem high, it should be remembered that we have assumed a value of $Z=1$, which is small. Recalling that the Born effective charges in CsNbW$_2$O$_9$ are anomalous, and using the relationship $P = Z\Gamma = \frac{1}{V}\sum_i q_iz_i$, where $q_i$ and $z_i$ are the effective charge and displacements of the $i$th atom, we estimated a value of $Z = 5.3$ for the proper phase. As $Z$ controls the strength of the coupling between the polar mode and the electric field, the coercive fields scale inversely with $Z$, so that they could be almost a magnitude smaller. At the very least, the values of the coercive fields in Figure \ref{electric} represent an upper limit for a homogenous switching of domains; the coercive fields could be even smaller if the typical switching mechanism involving nucleation and growth of domains is used.

The global stability of the proper-like minimum in CsNbW$_2$O$_9$ is intriguing, but the relative stability of the two minima is strongly dependent on simulation parameters. Indeed, if we use a supercell approach instead of the VCA (details in Tables S7 and S8), we observe that the improper minimum is just lower in energy. The pertinent observation is that the proper-like minimum exists with all computational methodologies we have used. To actually observe the proper minima in CsNbW$_2$O$_9$, we propose that a high quality single crystal be grown, or an ultrathin film so that electrical measurements such as Figure \ref{electric} can be performed, potentially at elevated temperatures to help deal with kinematical trapping in local minima. We note that the proper-like minimum does appear to be genuine as RbNbW$_2$O$_9$ \cite{chang2008polar,mcnulty2020phase} has been synthesised with characteristic proper behaviour.

The transition between proper-like and improper-like minima, and the associated change in polarisation, suggests that CsNbW$_2$O$_9$ is a ferrielectric \cite{maisonneuve1997ferrielectric,fu2020unveiling}. These are materials containing oppositely oriented polar distortions on at least two sublattices which do not fully compensate each other, leaving a net polarisation. In fact, the switching behaviour of Figure \ref{electric} is exactly what would be expected for a ferrielectric consisting of two sublattices (see Figure S8). Again, a clear distinction must be made here between CsNbW$_2$O$_9$ and YMnO$_3$. While YMnO$_3$ is similar in having sublattices of uncompensated antiparallel dipoles, it only has two stable (improper) polarisation states and so is functionally only ferroelectric rather than being considered ferrielectric. With this identification made, we conclude by noting that differing patterns of dipole cancellation in ferrielectrics can lead to a multivalued polarisation, and properties of this kind have been proposed as the basic operating principle for novel neuromorphic computing, which requires a near-continuum of states. In a macroscopic sample of CsNbW$_2$O$_9$, this could be achieved by tuning the volume fraction $x$ of the material occupying the proper-like minimum against the fraction $(1-x)$ in the improper-like. The resulting polarisation of the whole material would then be a near-continuous function of the parameter $x$, which changes as the proper/improper-like domains grow. The transition from proper to improper is also likely to have important ramifications for the domain microstructure in these materials, and introduce further flexibility in controlling the functional properties of domain walls for the burgeoning field of domain wall nanoelectronics \cite{meier2022ferroelectric,catalan2012domain}.

\textit{Conclusion - } We have explored the energy landscape of Equation \ref{eq:landscape} for hexagonal YMnO$_3$ and the hexagonal tungsten bronze CsNbW$_2$O$_9$. The symmetry of each dictate that they are both described by the same free energy expression, but the different nature of the $\Gamma_2^-$ and $K_3$ modes leads to drastically different values for the coefficients. Specifically, CsNbW$_2$O$_9$ has a strongly negative $a_\Gamma$, which forms additional minima in the free energy displaying proper-like ferroelectric behaviour. We predict that a decrease in temperature may cause a transition from the improper-like to the proper-like but suggest that such a transition is likely to be kinetically limited. Instead, electric fields could be used to perform the switching, in which a triple hysteresis loop would be observed - akin to that of ferrielectrics. We suggest that the hexagonal tungsten bronzes are a family of materials with a simple free energy expansion that nevertheless allows for large enhancements of the polarisation under electric fields, driven by the competition between proper-like and improper-like phases. The near degeneracy of the two minima could also prove a useful playground for exploring unusual polar textures \cite{junquera2023topological,das2019observation}.  We have also highlighted the potential of materials with these properties in designing novel electric memory devices beyond the two-state paradigm, as well as how they could find applications in the emerging field of neuromorphic computing.

\textit{Conflicts of interest} - The authors declare no conflicts of interest.

\textit{Acknowledgements} - The authors express their gratitude to J.M. Gregg for helpful discussions related to this work. This work used the Hamilton HPC service at Durham University. CAMS and NCB acknowledge the Leverhulme Trust for a research project grant (Grant No. RPG-2020-206).

\bibliography{bibliography}

\clearpage
\def\supplementfilename{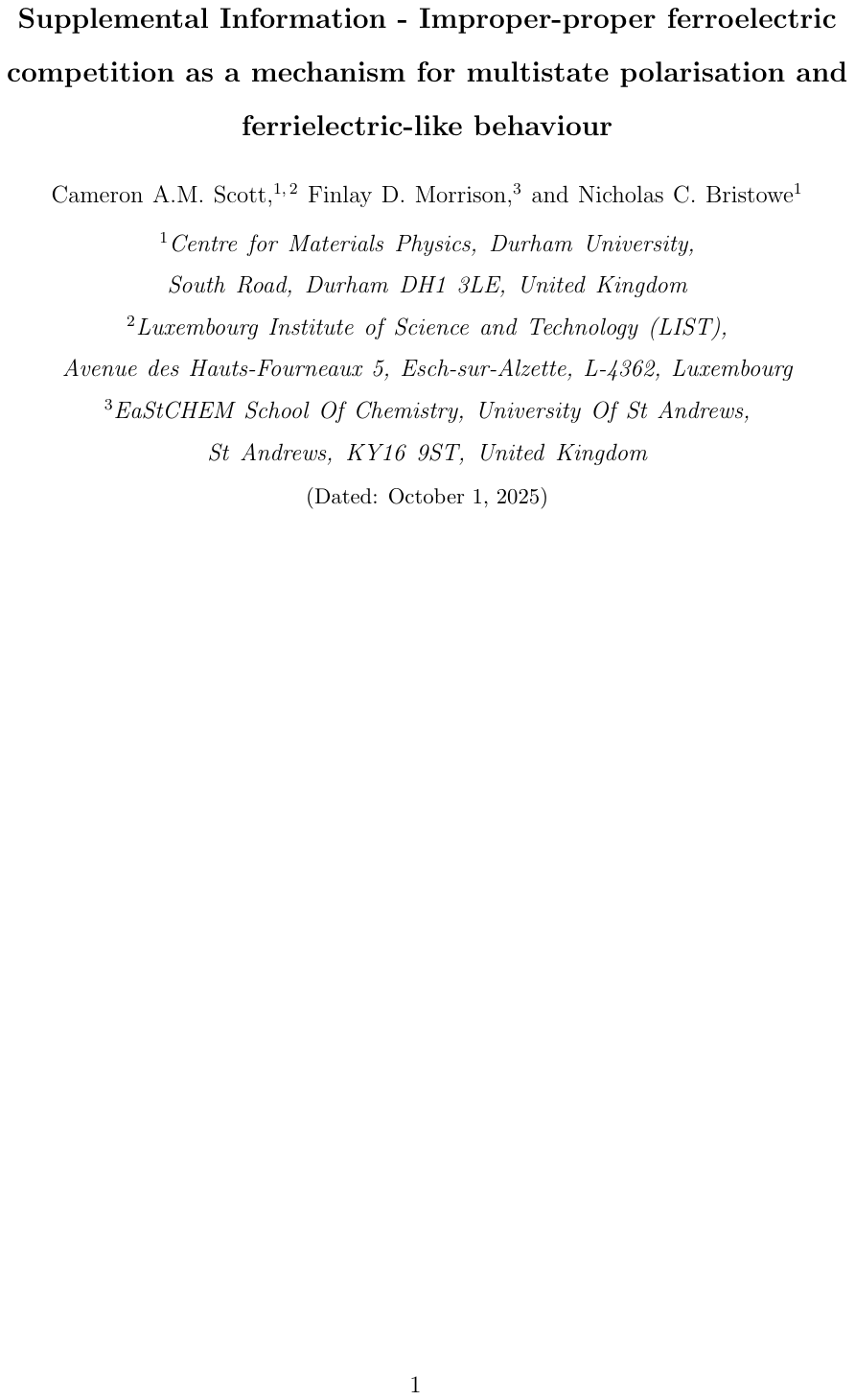}
\pdfximage{\supplementfilename}
\foreach \x in {1,...,\the\pdflastximagepages} {
  \clearpage
  

\section*{Supplementary material}

**DERIVATION OF DIELECTRIC SUSCEPTIBILITY**

To derive the dielectric constant from the free energy expansion

$$\mathcal{F}(\Gamma, K) = a_\Gamma \Gamma^2 + b_\Gamma \Gamma^4 + a_K K^2 + b_K K^4 + \beta \Gamma^2 K^2 + \lambda \Gamma K^3 \tag{1}$$

we take derivatives

$$\frac{\partial \mathcal{F}}{\partial \Gamma} = 2a_\Gamma \Gamma + 4b_\Gamma \Gamma^3 + 2\beta \Gamma K^2 + \lambda K^3 = 0, \tag{2}$$

and

$$\frac{\partial \mathcal{F}}{\partial K} = 2a_K K + 4b_K K^3 + 2\beta \Gamma^2 K + 3\lambda \Gamma K^2 = 0, \tag{3}$$

to find minima. Where both of these are true implicitly defines the locations of the minima in the free energy landscape $\Gamma_0(K)$ and $K_0(\Gamma)$ and are obtained numerically.

To find the dielectric susceptibility $\chi$, we add a coupling term $-EP = -EZ\Gamma$ to the free energy and use the definition $\chi = \frac{1}{\epsilon_0} \frac{\partial P_0}{\partial E} = \frac{Z}{\epsilon_0} \frac{\partial \Gamma_0}{\partial E}$, where $\epsilon_0$ is the permittivity of free space, to obtain

$$2a_\Gamma \chi + 12 b_\Gamma \Gamma_0^2 \chi + 2\beta \chi K_0^2 + 4\beta \Gamma_0 K_0 \chi \left.\frac{\partial K_0}{\partial \Gamma}\right|_{\Gamma=\Gamma_0} + 3\lambda K_0^2 \chi \left.\frac{\partial K_0}{\partial \Gamma}\right|_{\Gamma_0} = \frac{Z^2}{\epsilon_0} \tag{4}$$

which can be solved to give

$$\chi = \frac{1}{\epsilon_0} \frac{Z^2}{2a_\Gamma + 12 b_\Gamma \Gamma_0^2 + 2\beta K_0^2 + (4\beta \Gamma_0 K_0 + 3\lambda K_0^2) \left.\frac{\partial K_0}{\partial \Gamma}\right|_{\Gamma_0}}. \tag{5}$$

We require the derivative $\left.\frac{\partial K_0}{\partial \Gamma}\right|_{\Gamma_0}$ which can be obtained implicitly through Equation 3

$$\left.\frac{\partial K_0}{\partial \Gamma}\right|_{\Gamma=\Gamma_0} = -\frac{4\beta \Gamma_0 K_0 + 3\lambda K_0^2}{2a_K + 12 b_K K_0^2 + 2\beta \Gamma_0^2 + 6\lambda \Gamma_0 K_0}. \tag{6}$$

Equation 5, when substituted with Equation 6, can now be solved numerically.

TABLE S1. Energies of crystal structures after relaxation with the virtual crystal approximation

| Energy | $P6/mmm$ | $P6mm$ | $P6_3cm$ | $Cmc2_1$ |
|---|---|---|---|---|
| (meV/ f.u) | 0.000 | -0.595 | -0.605 | -0.639 |

2

TABLE S2. Structural properties of $CsNbW_2O_9$ without octahedral tilts. Parent and proper cells are enlarged to allow for the zone boundary modes. The structures are illustrated in Figure 2.

| Descriptor | $P6/mmm$ | $P6mm$ (Proper) | $P6mm$ (Improper) |
|---|---|---|---|
| $a$ (Å) | 12.743 | 12.586 | 12.568 |
| $b$ (Å) | 12.743 | 12.586 | 12.568 |
| $c$ (Å) | 3.803 | 4.082 | 4.084 |
| $V$ (Å$^3$) | 534.891 | 559.968 | 558.603 |
| $\Gamma_1^+$(a) (Å) | N/A | 0.002 | 0.008 |
| $\Gamma_2^-$(a) (Å) | N/A | 0.658 | 0.256 |
| $K_1$(a,0) (Å) | N/A | N/A | 0.007 |
| $K_3$(a,0) (Å) | N/A | N/A | 0.450 |
| $E$ (meV/f.u) | 0.000 | -597.602 | -574.943 |

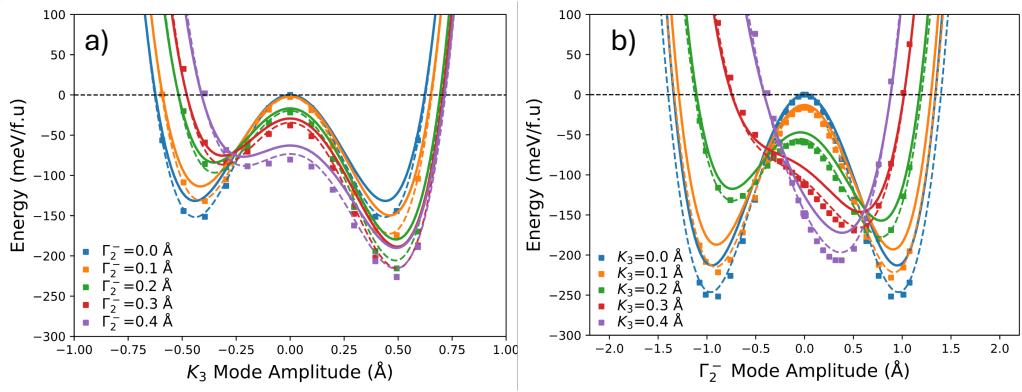

FIG. S1. Comparison of the energy landscape obtained by fitting the DFT data to a fourth-order energy polynomial (1) in solid lines, and with a fit to a sixth order polynomials in dashed lines. Fixing the fourth order coefficients to those in Table I, the sixth-order terms that are allowed by symmetry have the form $55\Gamma^6 - 21K^6 - 32\Gamma^3K^3 + 250\Gamma^2K^4 - 367\Gamma^4K^2 - 10\Gamma K^5$ where all the coefficients have units of m$e$V/ Å$^6$ f.u. .

3

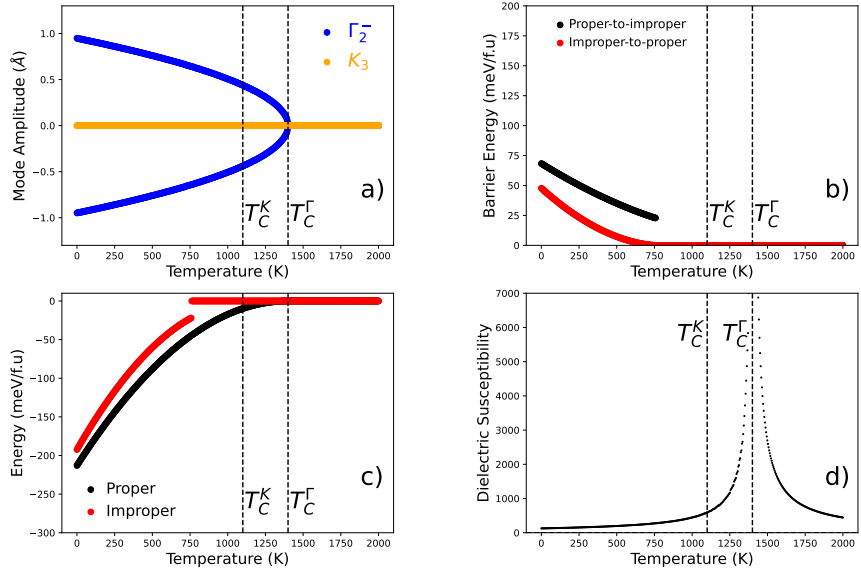

FIG. S2. Temperature dependence of a) order parameters, b) energy barriers, c) energies and d) dielectric susceptibility. Coefficients in the Landau model are : $a_\Gamma = -470$, $a_K = -1360$, $b_\Gamma = 260$, $b_K = 3490$, $\beta = 4280$ and $\lambda = -3380$. The two transition temperatures are set to $T_C^\Gamma = 1400$ K and $T_C^K = 1100$ K.

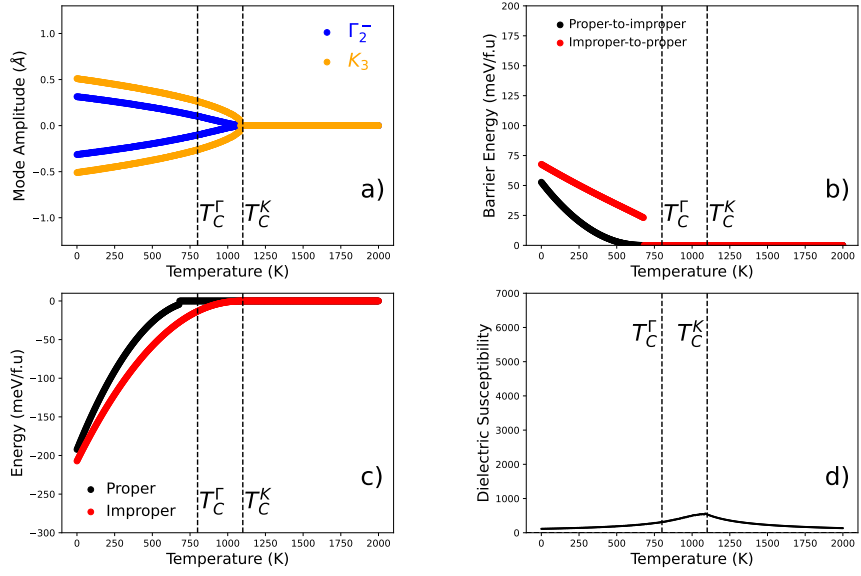

FIG. S3. Temperature dependence of a) order parameters, b) energy barriers, c) energies and d) dielectric susceptibility. Coefficients in the Landau model are : $a_\Gamma = -445$, $a_K = -1430$, $b_\Gamma = 260$, $b_K = 3490$, $\beta = 4280$ and $\lambda = -3380$. The two transition temperatures are set to $T_C^\Gamma = 800$ K and $T_C^K = 1100$ K.

5

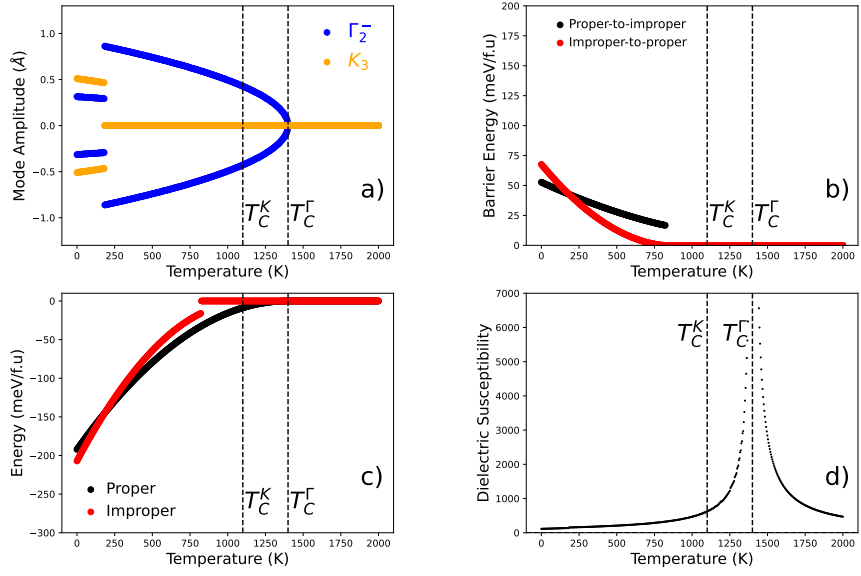

FIG. S4. Temperature dependence of a) order parameters, b) energy barriers, c) energies and d) dielectric susceptibility. Coefficients in the Landau model are : $a_\Gamma = -445$, $a_K = -1430$, $b_\Gamma = 260$, $b_K = 3490$, $\beta = 4280$ and $\lambda = -3380$. The two transition temperatures are set to $T_C^\Gamma = 1400$ K and $T_C^K = 1100$ K.

6

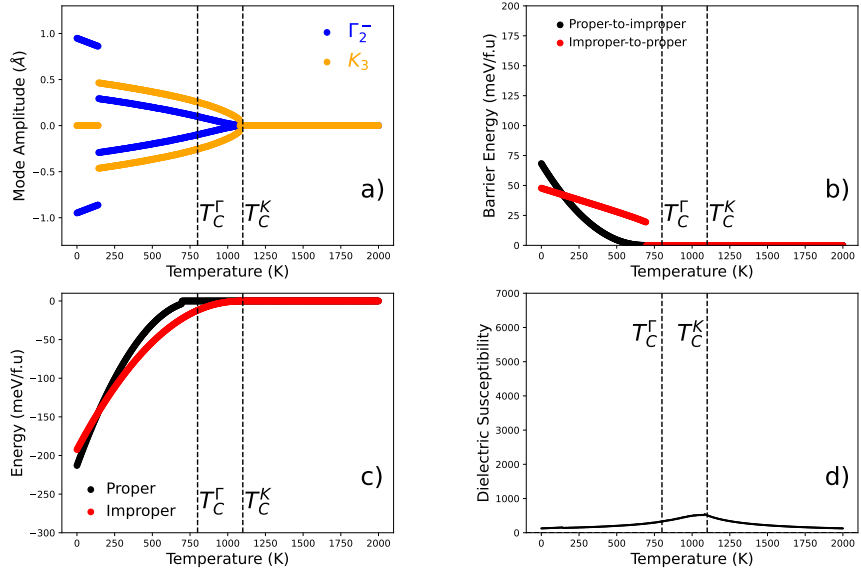

FIG. S5. Temperature dependence of a) order parameters, b) energy barriers, c) energies and d) dielectric susceptibility. Coefficients in the Landau model are : $a_\Gamma = -470$, $a_K = -1360$, $b_\Gamma = 260$, $b_K = 3490$, $\beta = 4280$ and $\lambda = -3380$. The two transition temperatures are set to $T_C^\Gamma = 800$ K and $T_C^K = 1100$ K.

7

TABLE S3. Fractional coordinates and site occupancies for the fully relaxed $P6/mmm$ structure. Lattice vectors are $a = b = 7.356$ Å and $c = 3.803$ Å, and the angles between lattice vectors are $\alpha = \beta = 90°$, $\gamma = 120°$.

| Atom | $x$ | $y$ | $z$ | Occupancy |
|---|---|---|---|---|
| Cs | 0.000 | 0.000 | 0.500 | 1 |
| Nb | 0.500 | 0.000 | 0.000 | $\frac{1}{3}$ |
| Nb | 0.000 | 0.500 | 0.000 | $\frac{1}{3}$ |
| Nb | 0.500 | 0.500 | 0.000 | $\frac{1}{3}$ |
| W | 0.500 | 0.000 | 0.000 | $\frac{2}{3}$ |
| W | 0.000 | 0.500 | 0.000 | $\frac{2}{3}$ |
| W | 0.500 | 0.500 | 0.000 | $\frac{2}{3}$ |
| O | 0.500 | 0.000 | 0.500 | 1 |
| O | 0.000 | 0.500 | 0.500 | 1 |
| O | 0.500 | 0.500 | 0.500 | 1 |
| O | 0.208 | 0.417 | 0.000 | 1 |
| O | 0.792 | 0.583 | 0.000 | 1 |
| O | 0.583 | 0.792 | 0.000 | 1 |
| O | 0.417 | 0.208 | 0.000 | 1 |
| O | 0.208 | 0.792 | 0.000 | 1 |
| O | 0.792 | 0.208 | 0.000 | 1 |

TABLE S4. Fractional coordinates and site occupancies for the fully relaxed $P6mm$ structure in the proper minimum. Lattice vectors are $a = b = 7.267$ Å and $c = 4.081$ Å, and the angles between lattice vectors are $\alpha = \beta = 90°$, $\gamma = 120°$.

| Atom | $x$ | $y$ | $z$ | Occupancy |
|---|---|---|---|---|
| Cs | 0.000 | 0.000 | 0.560 | 1 |
| Nb | 0.500 | 0.000 | 0.075 | $\frac{1}{3}$ |
| Nb | 0.000 | 0.500 | 0.075 | $\frac{1}{3}$ |
| Nb | 0.500 | 0.500 | 0.075 | $\frac{1}{3}$ |
| W | 0.500 | 0.000 | 0.075 | $\frac{2}{3}$ |
| W | 0.000 | 0.500 | 0.075 | $\frac{2}{3}$ |
| W | 0.500 | 0.500 | 0.075 | $\frac{2}{3}$ |
| O | 0.500 | 0.000 | 0.495 | 1 |
| O | 0.000 | 0.500 | 0.495 | 1 |
| O | 0.500 | 0.500 | 0.495 | 1 |
| O | 0.792 | 0.208 | 0.001 | 1 |
| O | 0.792 | 0.583 | 0.001 | 1 |
| O | 0.417 | 0.208 | 0.001 | 1 |
| O | 0.208 | 0.792 | 0.001 | 1 |
| O | 0.208 | 0.417 | 0.001 | 1 |
| O | 0.583 | 0.792 | 0.001 | 1 |

9

TABLE S5. Fractional coordinates and site occupancies for the fully relaxed $P6mm$ structure in the improper minimum. Lattice vectors are $a = b = 12.568$ Å and $c = 4.084$ Å, and the angles between lattice vectors are $\alpha = \beta = 90°$, $\gamma = 120°$.

| Atom | x | y | z | Occupancy |
|---|---|---|---|---|
| Cs | 0.000 | 0.000 | 0.583 | 1 |
| Cs | 0.333 | 0.667 | 0.539 | 1 |
| Cs | 0.667 | 0.333 | 0.539 | 1 |
| Nb | 0.833 | 0.167 | 0.110 | 1/3 |
| Nb | 0.833 | 0.667 | 0.110 | 1/3 |
| Nb | 0.333 | 0.167 | 0.110 | 1/3 |
| Nb | 0.167 | 0.833 | 0.110 | 1/3 |
| Nb | 0.167 | 0.333 | 0.110 | 1/3 |
| Nb | 0.667 | 0.833 | 0.110 | 1/3 |
| Nb | 0.500 | 0.000 | 0.968 | 1/3 |
| Nb | 0.000 | 0.500 | 0.968 | 1/3 |
| Nb | 0.500 | 0.500 | 0.968 | 1/3 |
| W | 0.833 | 0.167 | 0.110 | 2/3 |
| W | 0.833 | 0.667 | 0.110 | 2/3 |
| W | 0.333 | 0.167 | 0.110 | 2/3 |
| W | 0.167 | 0.833 | 0.110 | 2/3 |
| W | 0.167 | 0.333 | 0.110 | 2/3 |
| W | 0.667 | 0.833 | 0.110 | 2/3 |
| W | 0.500 | 0.000 | 0.968 | 2/3 |
| W | 0.000 | 0.500 | 0.968 | 2/3 |
| W | 0.500 | 0.500 | 0.968 | 2/3 |
| O | 0.833 | 0.167 | 0.531 | 1 |
| O | 0.833 | 0.667 | 0.531 | 1 |
| O | 0.333 | 0.167 | 0.531 | 1 |
| O | 0.167 | 0.833 | 0.531 | 1 |
| O | 0.167 | 0.333 | 0.531 | 1 |
| O | 0.667 | 0.833 | 0.531 | 1 |
| O | 0.500 | 0.000 | 0.546 | 1 |
| O | 0.000 | 0.500 | 0.546 | 1 |
| O | 0.500 | 0.500 | 0.546 | 1 |
| O | 0.792 | 0.000 | 0.035 | 1 |
| O | 0.000 | 0.792 | 0.035 | 1 |
| O | 0.208 | 0.208 | 0.035 | 1 |
| O | 0.208 | 0.000 | 0.035 | 1 |
| O | 0.000 | 0.208 | 0.035 | 1 |
| O | 0.792 | 0.792 | 0.035 | 1 |
| O | 0.125 | 0.667 | 0.039 | 1 |
| O | 0.333 | 0.459 | 0.039 | 1 |
| O | 0.541 | 0.874 | 0.039 | 1 |
| O | 0.874 | 0.333 | 0.039 | 1 |
| O | 0.666 | 0.541 | 0.039 | 1 |
| O | 0.459 | 0.125 | 0.039 | 1 |
| O | 0.333 | 0.874 | 0.039 | 1 |
| O | 0.541 | 0.667 | 0.039 | 1 |
| O | 0.125 | 0.459 | 0.039 | 1 |
| O | 0.667 | 0.125 | 0.039 | 1 |
| O | 0.459 | 0.333 | 0.039 | 1 |
| O | 0.874 | 0.541 | 0.039 | 1 |

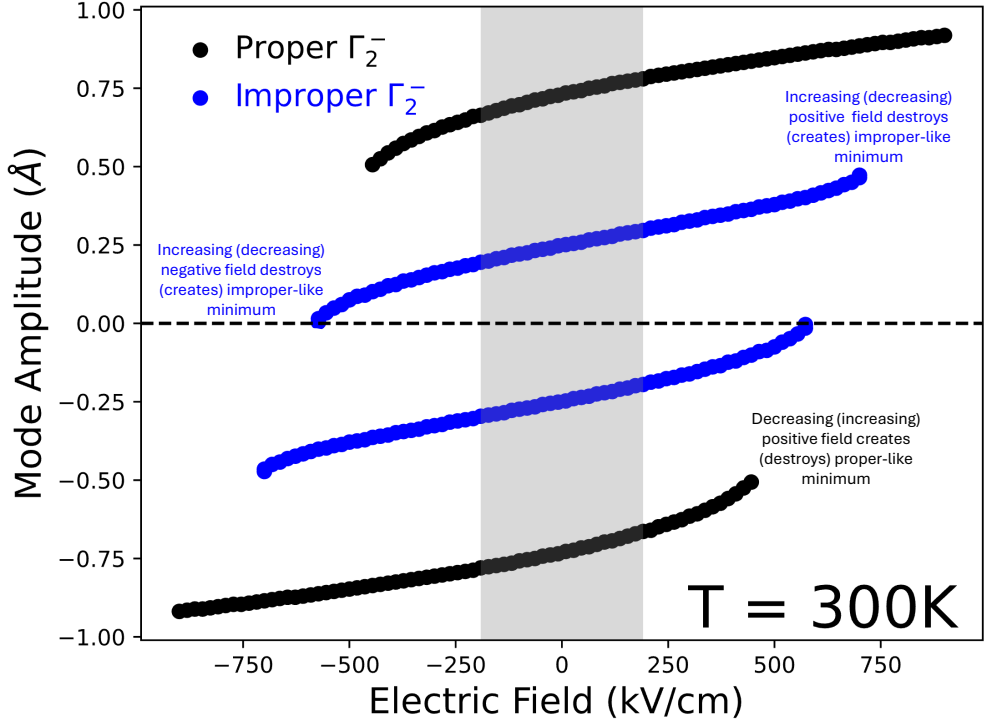

FIG. S6. The polar ($\Gamma_2^-$) mode amplitudes as a function of electric field at $T$=300 K. The shaded region denotes the fields for which the improper-like minima are lower energy. The applied electric field distorts the entire landscape of Figure 1d. It can create and destroy minima and change their relative energies. As the electric field does not directly couple to the antipolar $K_3$ mode, these modes have been omitted. In fact, the $K_3$ modes are only affected by electric fields through higher order couplings to $P$ and so the primary effect of an electric field is to shift the minima in Figure 1d horizontally.

11

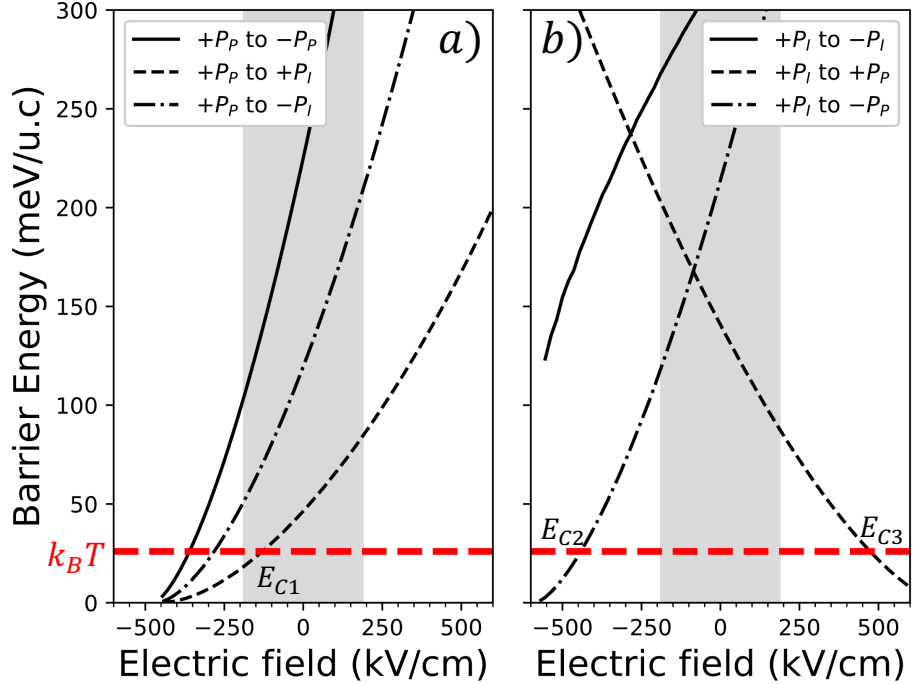

FIG. S7. The energy barrier between the various minima. Panel $a$) shows the energy barriers as a function of field for transition from the proper-like minimum with positive polarisation ($+P_P$) to the three other states the in free energy landscape of Figure 1. It is only possible to transition to the improper-like minima with polarisation $+P_I$ if those minima are lower energy than proper-like minima. The fields for which this is true are illustrated by the shaded region. There is only one energy barrier within this region that is less that the thermal energy $k_B$T, which gives the coercive field $E_{C1}$. Panel $b$) shows the energy barriers for transition away from the improper-like minima $+P_I$. A direct transition to the other improper-like minimum is not permitted without going over the central maximum of Figure 1d, and so only transitions to the proper-like minima are allowed but only outside the shaded region. This gives the other two coercive fields, $E_{C2}$ and $E_{C3}$.

12

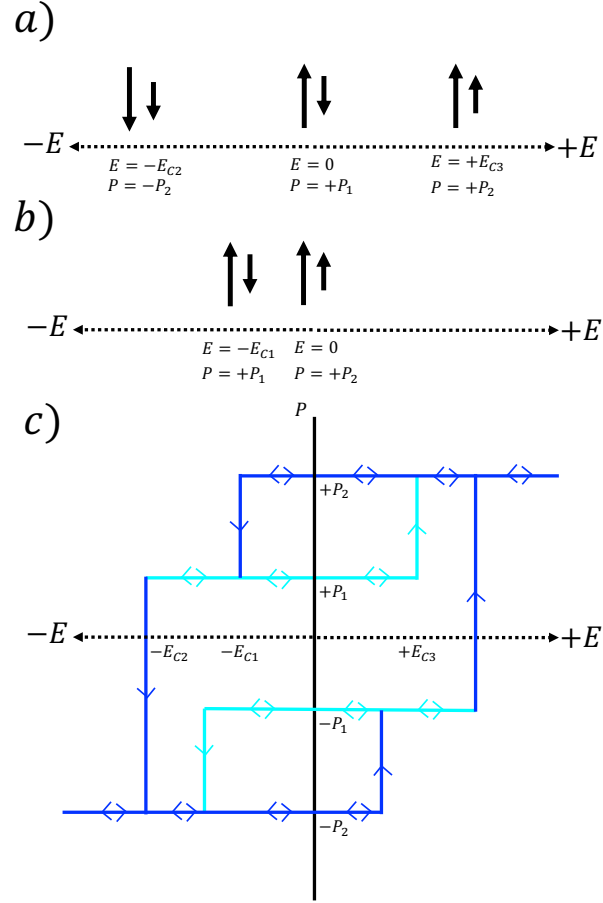

FIG. S8. Schematic cartoon displaying the switching behaviour of a ferrielectric consisting of two independent and unequal sublattices. $a)$ shows the behaviour under a field of the $+P_1$ state, whilst $b)$ shows the corresponding behaviour if the system is in the $+P_2$ state. The switching behaviour would be identical to that found in $CsNbW_2O_9$.

13

TABLE S6. Structural parameters of CsNbW$_2$O$_9$ including octahedral tilts. Parent and proper cells are enlarged to allow for the zone boundary modes.

| Descriptor | $P6/mmm$ | $Cmc2_1$ (Proper) | $Cmc2_1$ (Improper) |
|---|---|---|---|
| $a$ (Å) | 22.068 | 21.720 | 21.694 |
| $b$ (Å) | 12.741 | 12.661 | 12.603 |
| $c$ (Å) | 7.606 | 7.964 | 7.972 |
| $V$ (Å$^3$) | 2138.759 | 2190.117 | 2179.488 |
| $\Gamma_1^+$(a) (Å) | N/A | 0.027 | 0.024 |
| $\Gamma_5^+$(a,0) (Å) | N/A | 0.068 | 0.069 |
| $\Gamma_2^-$(a) (Å) | N/A | 1.077 | 0.658 |
| $\Gamma_5^-$(a,0.577a) (Å) | N/A | 0.036 | 0.025 |
| $A_3^+$(a) (Å) | N/A | 0.167 | 0.180 |
| $A_6^+$(a,0) (Å) | N/A | 0.772 | 0.828 |
| $A_4^-$(a) (Å) | N/A | 0.013 | 0.005 |
| $A_6^-$(0,a) (Å) | N/A | 0.180 | 0.150 |
| $H_2$(0,a) (Å) | N/A | N/A | 0.002 |
| $H_4$(0,a) (Å) | N/A | N/A | 0.015 |
| $H_5$(0,0,a,-a) (Å) | N/A | N/A | 0.075 |
| $H_6$(a,a,0,0) (Å) | N/A | N/A | 0.044 |
| $K_1$(a,0) (Å) | N/A | N/A | 0.005 |
| $K_3$(a,0) (Å) | N/A | N/A | 0.424 |
| $K_5$(0,0,a,-a) (Å) | N/A | N/A | 0.006 |
| $K_6$(a,-a,0,0) (Å) | N/A | N/A | 0.012 |
| $E$ (meV/f.u) | 0.000 | -648.900 | -638.558 |

TABLE S7. Comparison of the $P6/mmm$ symmetry of CsNbW$_2$O$_9$ using the VCA and a supercell (SC) method. An explicit ion ordering breaks the symmetry. We considered all cation orderings within a 2x2x2 supercell and found that the most common centrosymmetric symmetry is $P2/m$. We chose three of these at random to make SC1, SC2 and SC3.

| Descriptor | SC1 | SC2 | SC3 | SC Average | VCA | Difference |
|---|---|---|---|---|---|---|
| $a$ (Å) | 12.93 | 12.93 | 12.94 | 12.94 | 12.74 | -1.55% |
| $b$ (Å) | 12.91 | 12.93 | 12.92 | 12.92 | 12.74 | -1.39% |
| $c$ (Å) | 7.70 | 7.70 | 7.69 | 7.70 | 7.60 | -1.30% |
| $V$ (Å$^3$) | 1113.80 | 1114.05 | 1114.77 | 1114.21 | 1068.31 | -4.12% |
| $\alpha$ (°) | 90.00 | 90.00 | 90.00 | 90.00 | 90.00 | 0.00% |
| $\beta$ (°) | 119.97 | 120.04 | 119.94 | 119.98 | 120.00 | 0.02% |
| $\gamma$ (°) | 90.00 | 90.00 | 90.00 | 90.00 | 90.00 | 0.00% |

TABLE S8. Comparison of mode amplitudes and energies using VCA and supercell (SC) methods with the $\Gamma_2^-$ and $K_3$ modes added to SC1 and then a full geometry relaxation performed. The resulting structure was then decomposed into the symmetry adapted distortions. The energy $\Delta E$ refers to the energy difference between the parent structure (either $P6/mmm$ or $P2/m$) and the fully relaxed structures. The energy difference is smaller for the supercell approach because the cation order in the parent already permits several energy lowering distortions.

| Descriptor | VCA (Proper) | VCA (Improper) | SC1 (Proper) | SC1 (Improper) |
|---|---|---|---|---|
| $\Gamma_2^-$ (Å) | 0.658 | 0.256 | 0.549 | 0.306 |
| $K_3$ (Å) | N/A | 0.450 | 0.001 | 0.387 |
| $\Delta E$ (meV/f.u) | -597.602 | -574.943 | -164.236 | -189.827 |


}

\end{document}